\documentclass[twoside,fleqn,espcrc2]{article}
\usepackage{fleqn,espcrc2}
\usepackage{graphicx}
\usepackage{here}%pour inserer figures et tables a emplacement 

\newcommand{\AmS}{{\protect\the\textfont2
    A\kern-.1667em\lower.5ex\hbox{M}\kern-.125emS}}
\def\beq{\begin{equation}}
\def\eeq{\end{equation}}
\def\bea{\begin{eqnarray}}
\def\eea{\end{eqnarray}}
\def\bq{\begin{quote}}
\def\eq{\end{quote}}

\def\nnb{\nonumber}

\def\rar{\rightarrow}

\def\lrar{\leftrightarrow}

\def\nnb{\nonumber}
\def\la{\langle}
\def\ra{\rangle}
\def\nin{\noindent}
\def\ba{\begin{array}}
\def\ea{\end{array}}

\def\b{\bullet}

\title{\bf{\boldmath
{\Large Are the pentaquark sum rules reliable?\thanks{Some materials of this paper have been presented by
R. D. Matheus at the QCD 04 11th International Conference (Montpellier
5-9th July 2004), and by S. Narison at the HEP-MAD 04 2nd High-Energy
Physics International Conference (Antananarivo 27th Sept.-2nd Oct.
2004).}} }}
\author{
Ricardo D. Matheus \address{ Laboratoire de Physique Th\'eorique et
d'Astrophysique, Universit\'e de Montpellier II, Case 070, Place
Eug\`ene Bataillon, 34095 - Montpellier Cedex 05,
France. Email: matheus@axpfep1.if.usp.br}\thanks{Fellow of the S\~ao Paulo State Foundation for Research Support (FAPESP).}  and Stephan
Narison\address{ Laboratoire de Physique Th\'eorique et d'Astrophysique, Universit\'e
de Montpellier II, Case 070, Place Eug\`ene
Bataillon, 34095 - Montpellier Cedex 05,
France. E-mail:
snarison@yahoo.fr} }
\begin{document}
\pagestyle{plain}
%\vspace*{2cm}
%%%%%%%%%%%%%%%%%
\begin{abstract}
\noindent
We rewiew and scrutinize the existing mass determinations of the
pentaquarks from the exponential Laplace Sum Rules (LSR). We do not find
any {\it sum rule window} for extracting optimal and reliable
results from the LSR, due to the unusual slow convergence of the OPE
and to the exceptional important role of the QCD continuum into the
spectral function in this channel. Instead, we use in this channel, {\it for the first
time}, Finite Energy Sum Rules (FESR), which exhibit a nice stability in
the  QCD continuum threshold $t_c$, at which one can extract, with a good
accuracy, the mass of the lowest resonance. Including the $D=7,~9$
condensate contributions in the OPE, we obtain $M_\Theta=(1513\pm 114)$
MeV, and the corresponding residue
$\lambda_\Theta^2\approx -(0.14\sim 0.49)\times 10^{-9}$ GeV$^{12}$, which favours the $I=0$, $J=1/2$, and negative
parity S-wave interpretation of the
$\Theta (1540)$.  However, our analysis indicates a degeneracy between the unmixed $I=0$ and $I=1$ $S$-wave states. In the
$I=0$,
$J=1/2$, $P$-wave channel, we obtain, for the
$P$-resonance, $M_P \simeq  (1.99\pm 0.19)$ GeV and $\lambda_P\approx -(0.7\sim 7.1)\times 10^{-9}$ GeV$^{14}$,
which we expect to be discovered experimentally. Our results also suggest that
some {\it intuitive choices} of the continuum threshold used in the LSR literature are inconsistent with the FESR results. Finally, a study of
the
$\Theta$-$K$-$N$ coupling using a vertex sum rule shows that, for the
$I=0$, $S$-wave channel, the leading OPE contributions only start to order $\alpha_s^2$ in the chiral limit $m_s=0$, indicating that
the $\Theta$ is very narrow. 
\end{abstract}
\maketitle
%%%%%%%%%%%%%%%%%%%%%%%%%
\section{INTRODUCTION}
\nin
Recent experimental discovery of the $\Theta (1540)$ as a narrow KN bound state in $\gamma$-nucleon and
$\gamma$-nucleus processes, $e^+e^-$ and hadronic machines \cite{NICOLAI}\footnote{However,  this narrow state has
not been confirmed by several experiments with high statistics and very good particle identification, using either
$e^+e^-$
\cite{LISTA} and hadronic initial states.} have stimulated renewed theoretical interests in hadron spectroscopy
\cite{JAFFE,THEORY}.  In this paper, we shall critically re-analyze the mass determinations of the isoscalar $I=0$,
$\Theta$ pentaquark mass from the exponential Laplace sum rules (LSR) \cite{oka.04,zhu.03,makus.04} within the diquark
scenario \cite{JAFFE} and propose new analysis using Finite Energy Sum Rule (FESR).
%%%%%%%%%%%%%%%%%%%%%%%%%%%%%%%%%
\vspace*{-0.5cm}
\section{THE PENTAQUARK CURRENTS}
\nin
The basic ingredients in the resonance mass determinations from QCD spectral sum rules \cite{SVZ,SNB} as well as from
lattice QCD calculations are the choice of  the interpolating 
currents for describing the resonance states. Contrary to the ordinary mesons, where the form of the current comes
from first principles, there are different choices of the pentaquark currents in the literature. The following analysis
also postulates the existence of a strongly bound pentaquark resonance. 
We shall list below some possible operators describing the isoscalar
$I=0$ and
$J=1/2$ channel\footnote{The isovector 
$I=1$ current for $S$-wave
resonance have been proposed by \cite{mat.02}. We have also checked (see also \cite{NIELSEN}) that the tensor diquark current used in
\cite{IOFFE} is an isovector $I=1$ instead of an isoscalar $I=0$ as confirmed by
the authors  in the revised version of their paper. In the following, we shall neglect the isospin breaking discussed in
\cite{VENEZIA}, which would come from higher order diagrams in our analysis.}, which would correspond to the experimental
candidate
$\Theta(1540)$ \cite{NICOLAI}. 
%%%%%%%%
%\subsection*{The $\Theta(1.54)$ interpolating currents}
\nin
Defining the pseudoscalar ({\it ps}) and scalar ({\it s}) diquark interpolating fields as:
\bea
\label{RD1.pseudo}
Q^{ps}_{ab}(x) &=& \left[u_{a}^{T}(x)Cd_{b}(x)\right]~,\nnb\\
Q^{s}_{ab}(x) &=& \left[u_{a}^{T}(x)C\gamma_{5}d_{b}(x)\right]~,
\eea
where $a,~b,~c$ are colour indices and $C$ denotes the charge conjugation matrix, 
 the lowest dimension current built by two diquarks and one anti-quark describing the $\Theta$ as
a $I=0,~J^P=1/2^+$ $S$-wave resonance is \cite{oka.04} (see also \cite{SASAKI})\footnote{A negative parity state can be obtained
by multiplying by $\gamma_5$ the diquark operator.}:
\begin{equation}
\label{RD1T.oka}
    \eta^{\Theta}_{\mbox{{\small \cite{oka.04}}}} = \epsilon^{abc} \epsilon^{def} \epsilon^{cfg}  Q_{ab}^{ps} 
   Q_{de}^{s}  C \bar{s}_{g}^{T}~,
\end{equation} 
and the one with one diquark and three quarks is~\cite{zhu.03}:
\begin{equation}
\label{RD1.zhu}
  \eta^{\Theta}_{\mbox{{\small \cite{zhu.03}}}} = \frac{1}{\sqrt{2}}\epsilon^{abc}  
  Q^{s}_{ab} \left\{ u_{e} \bar{s}_e i \gamma_5 d_c - (u \leftrightarrow d) \right\}~. 
\end{equation} 
This later choice can be interesting if the instanton repulsive force arguments \cite{SHURYAK} against the existence of a pseudoscalar
diquark bound state  apply. Alternatively, a description of the $\Theta(1.54)$ as a $I=0$, $J^P=1/2^+$ $P$-wave resonance has been
proposed by
\cite{JAFFE} and used by
\cite{makus.04} in the sum rule analysis:
\bea
\label{RD1.eidcur}
  \eta^\Theta_{\mbox{{\small \cite{makus.04}}}} = \left( \epsilon^{abd} \delta^{ce} +  \epsilon^{abc} \delta^{de}
\right) 
  [Q^{s}_{ab}(D^{\mu}Q^{s}_{cd})-\nnb\\
(D^{\mu}Q^{s}_{ab})Q^{s}_{cd}] \gamma_{5} \gamma_{\mu} C \bar{s}_{e}^{T}~.
\eea
We have generalized this current by considering its mixing with the following one having the same dimension and quantum numbers:
\label{RD2.eidcur}
\begin{equation}\label{eta.new}
  \eta^\Theta_{{{\small new}}} = \epsilon^{abc} \epsilon^{def}  \epsilon^{cfg}
 Q^{ps}_{ab}Q^{s}_{de}\gamma_\mu (D^{\mu} C \bar{s}_{g}^{T})~.
\end{equation}
%%%%%%%%%%
%\subsection*{The $\Xi(1.86)$ interpolating currents}
%The two diquarks and one anti-quark current decribing the $\Xi(1860)$ as a $I=0,~J^P=3/2^+$ $S$-wave resonance is \cite{oka.04}:
%\begin{equation}
%\label{RD1C.oka}
%    \eta^{\Xi}_{\mbox{{\small \cite{oka.04}}}} = \epsilon^{abc} \epsilon^{def} \epsilon^{cfg}  \qs_{ab}^{ps} 
%   \qs_{de}^{s}  C \bar{u}_{g}^{T}~.
% \end{equation}
%%%%%%%%%%%%%%%%%%%%%%%%%%%%%%%%%%%%%
\section{THE QCD SPECTRAL FUNCTIONS}
\nin
For the QCD spectral sum rules analysis,
we shall work here with the
two-point correlators: 
\bea
\Pi^H(q^2) &\equiv& i \int d^4x ~e^{iqx} \
\la 0\vert {\cal T}
\eta^H(x)
\bar{\eta}^H(0) \vert 0 \ra ,
\eea
built from the previous $\eta$ currents. It possesses the Lorentz decomposition:
\beq
\label{eq: invariant}
\Pi^H(q^2)= \hat q A^H(q^2)+ B^H(q^2)~.
\eeq
$\b$ The QCD expression of the correlators associated to different choices of the currents
is known in the literature \cite{oka.04,zhu.03,makus.04} to leading order of PT series and including
the three first non-perturbative condensate ($D\leq 5,6$) contributions. We have checked the QCD expressions given there and agree
with their results. However, at that approximation, we have added some missing contributions in \cite{zhu.03}.\\
$\b$ We have not
included in the OPE the contribution of the $D=2$ tachyonic gluon mass induced by the resummation of the PT series \cite{ZAK}
bearing in mind that this effect will be negligible, to the accuracy we are working, as illustrated in some examples
\cite{TACH}. \\
$\b$ We have included the $D=7,9$ contributions into the QCD expression of the spectral function associated to the current 
in Eq. (\ref{RD1T.oka}), which we shall extensively study as
a prototype example in this paper. In doing this calculation, we have worked in the chiral limit $m_s=0$, such that for consistencies, we
shall use the $SU(3)$ symmetric value $\la\bar ss\ra=\la\bar dd\ra$ for these contributions. In this particular example,
we have checked that the contribution of the four-quark condensate vanishes to leading order. In our preliminary results,
we also found that its radiative correction though not identically zero gives a negligible contribution. 
 We have also neglected the
contributions of the three-gluon condensate of the type g$\la GGG\ra\la\bar ss\ra$ assuming that the theorem in \cite{MALLIK} for
the light quark bilinear operators continues to hold for the diquark correlators \footnote{We plan to check explicitly this
result in a future publication.}, which factorize during the evaluation of the QCD expression. \\
$\b$ We have evaluated the new contributions associated to the current 
$\eta_{new}$  in Eq. (\ref{eta.new}), where we found that, to leading order in $\alpha_s$ and in the
chiral limit
$m_q\rar 0$, the contribution to the correlator vanishes. This result justifies a posteriori the
{\it unique choice} of operator for the $P$-wave state used in \cite{JAFFE,makus.04}.
%%%%%%%%%%%%%%%%%%%%%%%%%%%%%%%%%%%%%
\section{THE LAPLACE SUM RULES (LSR)}
\nin 
We shall be concerned with the Laplace transform sum rules:
\bea
\label{eq:lapl}
{\cal L}^H_{A/B}(\tau)
&\equiv& \int_{t_\leq}^{\infty} {dt}~\mbox{e}^{-t\tau}
~\frac{1}{\pi}~\mbox{Im}{A^H/B^H}(t),\nnb \\ 
{\cal R}^H_{A/B}(\tau) &\equiv& -\frac{d}{d\tau} \log {{\cal L}^H_{A/B}(\tau)},
\eea
where $t_\leq$ is the hadronic threshold, and H denotes the corresponding hadron. The latter sum  rule,
 or its slight modification, is useful, as it is equal to the 
resonance mass squared, in  
 the usual duality ansatz parametrization of the spectral function:
\bea
\frac{1}{\pi}\mbox{ Im}A^H/B^H(t)\simeq (\lambda^2_H/\lambda^2_HM_H)\delta(t-M^2_H)
 \ + \ \nnb\\
 ``\mbox{QCD continuum}" \Theta (t-t_c),
\eea
where the ``QCD continuum comes from the discontinuity of the QCD
diagrams, which is expected to give a good smearing of the
different radial excitations . $\lambda_H$ is
the residue of the hadron $H$;
$t_c$ is the QCD continuum threshold, which is, like the 
sum rule variable $\tau$, an  (a priori) arbitrary 
parameter. In this
paper, we shall look for the
$\tau$- and $t_c$-stability criteria for extracting the optimal
results. 
For illustrating our analysis, we
give below the checked and completed LSR of the $S$-wave current in Eq. (\ref{RD1T.oka}) including the new $D=7,9$
high-dimension condensates in $B$:
\begin{eqnarray}
\label{eq: lsra}
&&{\cal L}^\Theta_{A_{\cite{oka.04}}}(\tau)=\frac{\tau^{-6}E_5}{860160\pi^{8}}+
\frac{\tau^{-4}E_3}{30720\pi^{6}}m_s\la\bar{s}s\ra+\nnb\\
&&\frac{\tau^{-4}E_3}{122880\pi^{7}}\la{\alpha_s}G^2\ra
-\frac{\tau^{-3}E_2}{36864\pi^{6}}m_sg\la\bar{s}{\bf\sigma G}s\ra~,
\end{eqnarray}
\vspace{-0.35cm}
\begin{eqnarray}
\label{eq: lsrb}
&&{\cal L}^\Theta_{B_{\cite{oka.04}}}(\tau)=\frac{\tau^{-6}E_5}{122880\pi^{8}}m_s
-\frac{\tau^{-5}E_4}{15360\pi^{6}}\la\bar{s}s\ra+,
\nnb\\
&&
\frac{\tau^{-4}E_3}{12288\pi^{6}}g\la
\bar{s}{\bf\sigma.G}s\ra+\frac{\tau^{-4}E_3}{24576\pi^{7}}m_s\la{\alpha_s}G^2\ra-\nnb\\
&&\frac{\tau^{-3}7E_2}{27648\pi^{5}}\la\bar{s}s\ra\la{\alpha_s}G^2\ra+\frac{\tau^{-2}E_1}{6144\pi^{4}}\la\bar{s}s\ra g\la
\bar{s}{\bf\sigma.G}s\ra ~,\nnb\\
\end{eqnarray}
\vspace{-0.2cm}
where:
\vspace{-0.1cm}
\beq
E_n=1-\Big{[}\rho_n\equiv e^{-t_c\tau}\sum_{0}^{n}\frac{(t_c\tau)^k}{k!}\Big{]}~,
\eeq
\vspace{-0.2cm}
$\rho_n$ being the notation in \cite{SNB}, while:
$\la\bar{s}s\ra,~\la\alpha_s G^2\ra$ are respectively the dimension $D=3$ quark and $D=4$
gluon condensates;
$g\la
\bar{s}{\bf\sigma G}s\ra\equiv g\la\bar{s}\sigma^{\mu\nu}(\lambda_a/2) G_{\mu\nu}^as\ra\equiv M_0^2\la \bar ss\ra$
is the $D=5$ mixed condensate. Throughout this paper we
shall use the values of the QCD parameters given in Table 1. 
\vspace{-1.cm}
\begin{table}[H]
\begin{center}
% space before first and after last column: 1.5pc
% space between columns: 3.0pc (twice the above)
\setlength{\tabcolsep}{.28pc}
% -----------------------------------------------------
% adapted from TeX book, p. 241
\newlength{\digitwidth} \settowidth{\digitwidth}{\rm 1.5}
%\catcode`?=\active \def?{\kern\digitwidth}
% -----------------------------------------------------
\caption{\footnotesize QCD input parameters used in the analysis.}
%\begin{tabular*}{\textwidth}{@{}l@{\extracolsep{\fill}}rrrrr}
\begin{tabular}{ll}
%\\
%\\
\hline 
%\hline
% \\
Parameters&References \\
%\\
\hline 
%\\
$\bar m_s(2~{\rm GeV})= (111\pm 22)$ MeV&\cite{SNB,QMASS,SNMS,PDG}\\
$\la \bar dd\ra^{1/3}$(2 GeV)=$-(243\pm 14)$ MeV&\cite{SNB,QMASS,DOSCHSN}\\
$\la \bar ss\ra /\la \bar dd\ra=0.8\pm 0.1$&\cite{SNB,QMASS,SNP2}\\
$\la \alpha_s G^2\ra=(0.07\pm 0.01)$ GeV$^4$&\cite{SNB,SNG}\\
$M^2_0=(0.8\pm 0.1)$ GeV$^2$&\cite{SNB,SNSP}\\
%&\\
%\\
\hline 
%\hline
\end{tabular}
\end{center}
\end{table}
\vspace{-1.cm}
\nin
We study the LSR in Eqs.
(\ref{eq: lsra}) and (\ref{eq: lsrb}). We find that all LSR corresponding to different currents present
the common features shown in Fig. \ref{fig: lsr}:
%%%%%%%%%%%%%%%%%%%%
%\vspace{-.5cm}
\begin{figure}[hbt]
\begin{center}
\includegraphics[width=6cm]{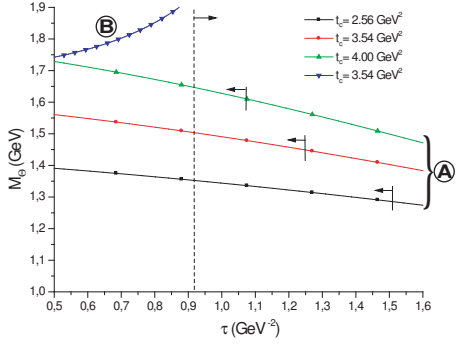}
\vspace{-.5cm}
\caption{\footnotesize $\tau$-behaviour of $M_\Theta$ for given values of  $t_c$. On the LHS of the vertical dashed line, the OPE
converges. The vertical line with arrow on each curve shows that the continuum contribution dominates over the resonance in the LHS
region. A (resp B) corresponds to the invariant defined in Eq. (\ref{eq: invariant})} 
\label{fig: lsr}
\vspace{-1.3cm}
\end{center}
\end{figure}
%%%%%%%%%%%%%%%%%%%%
\begin{enumerate}
\vspace{-.2cm}
\item[$-$]The $B$-component increases rapidly with $\tau$. Then, it is useless at that approximation of the OPE.
\vspace{-.5cm}
\item[$-$] For $A$, the mass prediction decreases smoothly when $\tau$ increases. The OPE converges for
$\tau\leq 0.9$ GeV$^{-2}$ (LHS of the vertical dashed line).\vspace{-.3cm}
\item[$-$] The QCD continuum contribution dominates over the resonance one in all ranges of $\tau$ where the OPE converges (LHS of the 
vertical line with arrow on each curve).
Indeed, for $\tau\leq 0.9$ GeV$^{-2}$, the QCD continuum contribution eats more than 84\% of the OPE one.
\end{enumerate}
\vspace{-.2cm}
$\b$ Therefore, it is impossible to find a {\it sum rule window} region where both the
resonance dominates over the QCD continuum contribution, and where the OPE converges.
Intuitively, this feature is expected as the current describing the pentaquark is of
higher dimensions, and therefore is more affected by  the continuum contribution than
the well-known sum rule for ordinary $\bar qq$ mesons. The absence of the sum rule
window is reflected by the increase of the mass predictions with the QCD continuum
threshold
$t_c$. In existing literature, the $t_c$-values have been fixed ad hoc and intuitively. 
\\
$\b$ During the evaluation of the different QCD diagrams, we do not find (to leading order in $\alpha_s$) any
factorization of the $(\bar s u)$-$(udd)$ diagram corresponding to a reducible $K$-$N$ continuum diagram, which has nothing to do
with the diquark picture. Then, our direct observation does not support the criticisms raised in \cite{KONDO.04} and refuted in \cite{LEE.04} on
a possible double counting due to the non-subtraction of the reducible diagram in the existing sum rules analysis of the $\Theta$. 
\\
$\b$  We conclude from the previous prototype example that the LSR using the simple duality ansatz: resonance+QCD continuum 
criterion is not appropriate for determining the pentaquark masses due to the absence of 
the usual {\it sum rule window}. Due to the huge continuum contribution ($\approx
85\%$) at relatively large $\tau\approx 1$ GeV$^{-2}$, the LSR
cannot strictly indicates the existence of the resonance into the spectral function.\\
$\b$ We have 
checked (though not explicitly shown in the paper) that the conclusions reached in the paper also apply to the sum rules used in the literature:
\cite{zhu.03} (current in Eq. (\ref{RD1.zhu})) and susbsequent uses in \cite{KONDO.04,LEE.04} for the $I=0$,
$S$-wave state; the sum rules used in \cite{mat.02} for the $I=1$, $S$-wave state; in \cite{makus.04} 
(current in Eq. (\ref{RD1.eidcur})) for the
$I=0,~1$
$P$-wave state; in \cite{IOFFE} for the $I=1$ tensor current and the
sum rules used in \cite{NAVARRA,KONDO.14} for studying the $J=3/2$ states. Indeed, in most LSR, the OPE does not converge at the
scale where the results are extracted, while the QCD continuum threshold has been taken arbitrarily or intuitively.\\
$\b$ The above results raise some doubts
on the validity of the results obtained so far in the existing literatures. Indeed, if one insists on using the LSR for predicting the
$\Theta$ parameters and some other pentaquark states, it is mandatory to introduce a more involved parametrization of the continuum spectral
function.
%%%%%%%%%%%%%%%%%%%%%%%%%%%%%%%%%%
\section{FINITE ENERGY SUM RULES (FESR)}
\nin
In contrast to the LSR, Finite Energy Sum Rules (FESR) \cite{RAF,KRAS,SNB} have the advantage to project out a set of
constraints for operators of given dimensions (local duality). They also correlate the resonance mass and residue to
the QCD continuum threshold $t_c$, so avoiding inconsistencies of the values of these parameters. Also
contrary to the LSR, the resonance and QCD continuum contributions are separated from the very beginning. The
FESR read:
\vspace{-0.2cm}
\bea
{\cal M}^H_{n}(A/B) &\equiv& \int_{t\leq}^{t_c}{dt}~t^n\mbox{
Im}A^H/B^H|_{EXP}\nnb\\
&\simeq &\int_{t\leq}^{t_c}{dt~t^n}\mbox{ Im}A^H/B^H|_{QCD}~.
\eea
 From the expressions of the spectral function given
previously, one can easily derive the FESR constraints.
\\
Doing the FESR analysis for the $A(q^2)$ invariant, one can notice that, at the approximation where the OPE is known ($D\leq 6$),
one does not have a stability in
$t_c$ for different moments ${\cal M}^\Theta_{n}$ and for different choices of the currents (see Fig. \ref{fig:
mass1}). Therefore, we will not consider this invariant in the paper.
%%%%%%%%%%%%%%%%%%%%%%%%%%%%%%%%%%%%%%%%
\subsection*{The $I=0$, $S$-wave channel}
\begin{figure}[hbt]
\begin{center}
\includegraphics[width=6cm]{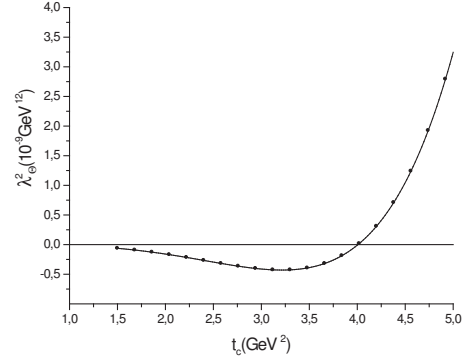}
\vspace{-.5cm}
\caption{\footnotesize $t_c$-behaviour of $\lambda^2_\Theta\times 10^9$ in GeV$^9$ including $D\leq 5$ condensates
in the OPE. } 
\label{fig: okalambda}
\vspace{-1.1cm}
\end{center}
\end{figure}

\nin
We illustrate the analysis by the current in Ref. \cite{oka.04} (the other choice \cite{zhu.03} in Eq.
(\ref{RD1.zhu})) has approximately the same dynamics as one can inspect from the QCD expressions). Including
the $D=7$ and 9 condensate contributions, the two first lowest dimension constraints from the
$B(q^2)$ invariant read:
\bea
{\cal M}^\Theta_{0,\cite{oka.04}} &=& {\frac {{\it m_s}\,{{\it t_c}}^{6}}{{88473600}{\pi }^{8}}
}-{\frac {{\la\bar{s}s\ra}\,{{ t_c}}^{5}}{1843200\pi ^{6}}}\nnb\\
&+&
\,{\frac {{\it g\la\bar{s}{\bf\sigma.G}s\ra}\,{{\it t_c}}^{4}}{294912{\pi }^{6}}
}
+{{m_s\la\alpha_s G^2\ra t_c^4}\over{589824\pi^7}}
\nnb\\
&-&{7\la\alpha_s G^2\ra\la\bar ss\ra t_c^3\over 165888\pi^5}+{\la\alpha_s G^2\ra{\it g\la\bar{s}{\bf\sigma.G}s\ra}t_c^2\over 12288\pi^5}
\eea
\bea
{\cal M}^\Theta_{1,\cite{oka.04}} &=& {\frac {{ m_s}\,{{ t_c}}^{7}}{{103219200}{\pi }^{8}}
}-{\frac {{\la\bar{s}s\ra}\,{{ t_c}}^{6}}{2211840\pi ^{6}}}\nnb\\
&+&
\,{\frac {{ g\la\bar{s}{\bf\sigma.G}s\ra}\,{{ t_c}}^{5}}{368640{\pi }^{6}}
}
+{{m_s\la\alpha_s G^2\ra t_c^5}\over{737280\pi^7}}
\nnb\\
&-&{7\la\alpha_s G^2\ra\la\bar ss\ra t_c^4\over 221184\pi^5}+{\la\alpha_s G^2\ra{\it g\la\bar{s}{\bf\sigma.G}s\ra}t_c^3\over 18432\pi^5}
\eea
%where:
%\bea
%{\cal I}(t_c,t_0)&=&1+{{\rm GeV}^4\over t_c^2}\Big{[}{2}\log{t_0\over t_c}+{159\over 5}\Big{]}~,\nnb\\
%{\cal I}(t_c,t_0)&=&1+{{\rm GeV}^4\over t_c^2}\Big{[}{5\over 3}\log{t_0\over t_c}+{85\over 12}\Big{]}~,
%\eea
%comes from loop-integrations.
from which one can deduce the mass squared:
\begin{equation}
M^2_{\Theta}\simeq {\frac{{\cal M}^\Theta_{1,\cite{oka.04}}}{{\cal M}^\Theta_{0,\cite{oka.04}}}}~.
\end{equation}
%%%%%%%%%%%%%%%%%%%%
%\vspace{-.5cm}
\begin{figure}[hbt]
\begin{center}
\includegraphics[width=6cm]{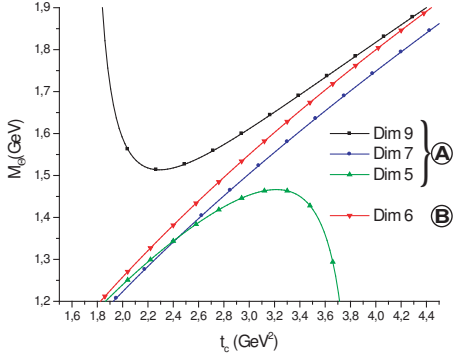}
\vspace{-.5cm}
\caption{\footnotesize $t_c$-behaviour of $M_\Theta$ for different truncations of the OPE. } 
\vspace{-1.cm}
\label{fig: mass1}
\end{center}
\end{figure}
%%%%%%%%%%%%%%%%%%%%
The behaviour of $M_{\Theta}$ is shown in Fig. \ref{fig: mass1}, for different truncations of the OPE.  One can
notice a stability at
$
{t_c}\simeq 2.29~ {\rm GeV}^2~, 
$
where the OPE starts to converge after only the inclusion of the $D=7+9$ condensates, while $D=7$ alone destroys the
stability reached for $D\leq 5$.
One can notice the important contribution of the lowest quark and mixed quark-gluon condensates in the OPE, which play a crucial
role in this mass determination. To that order, we obtain:
\vspace{-0.2cm}
\bea\label{eq:smass}
M_\Theta&\simeq& (1513\pm 20\pm 10\pm 40 \pm 30 \pm 30\pm 95)~{\rm GeV}\nnb\\
&\simeq&(1513\pm 114)~{\rm GeV}~,
\eea
%\vspace{-0.2cm}
where the errors come respectively from $m_s,~\la\bar qq\ra,~\la\alpha_s G^2\ra,~M^2_0$, the estimate of the higher dimension
condensates and the violation of the vacuum saturation assumption of the $D=7,9$ condensates by a factor $(2\pm 1)$
like in the
$\rho$-meson
\cite{TARRA,RAF} and some other channels \cite{SNB}. 
One can notice that:\\
$\b$ The existence of the $t_c$-stability point makes the superiority of FESR compared to the LSR in this channel. For the LSR,
$M_\Theta$ increases with $t_c$. Here, the
localisation of the stability point induces here a negligible error. \\
$\b$ The FESR order parameter in the OPE, $t_c\simeq 2.3$ GeV$^2$ is much larger
than for the LSR ($\tau^{-1}\leq 1$ GeV$^2$), implying a much better convergence of the OPE for the FESR,
and then a much more reliable result than the LSR.\\
$\b$ Working with ratio of higher moments $ {\cal M}^\Theta_{2}/{\cal M}^\Theta_{1}$,...leads to
almost the same value of $M_\Theta$. The slight variation is much smaller than the error in Eq. (\ref{eq:smass}).\\
$\b$ Truncating the OPE at $D=5$ like done in the available literature would give a slightly lower value of $M_\Theta$  at the
stability point $t_c\approx 3.2$ GeV$^2$ (see Fig. \ref{fig: mass1}), but compatible
with the one in Eq. (\ref{eq:smass}). 
\\ 
$\b$ Contrary to $M_\Theta$, the value and the sign of $\lambda_\Theta^2$ are very sensitive to the truncation of the OPE due to the
alternate signs of the condensate contributions in the analysis. The stability in $t_c$ is obtained after the inclusion of the $D=5$
condensate contributions as shown in Fig. \ref{fig: okalambda}. 
%%%%%%%%%%%%%%%%%%%%
%\vspace{-.4cm}
%%%%%%%%%%%%%%%%%%%%
To our approximation $D\leq 9$, the most conservative result is:
\beq
\lambda_\Theta^2\approx -(0.14\sim 0.49)\times 10^{-9}~{\rm GeV}^{12},
\eeq
where the range comes from the shift of the $t_c$-stability point from $D=5$ to $D=9$ approximation.
This result, though inaccurate, suggests that the parity of the
$\Theta$ is negative, as indicated by the lattice results given in
\cite{LATT}\footnote{At the approximation $D\leq 5$ 
the LSR does not converge such that analogous results obtained in \cite{oka.04,KONDO.04,LEE.04} should be taken with a great care.}.
Improving the accuracy of our result requires more high-dimension condensate terms in the OPE.
%%%%%%%%%%%%%%%%%%%%%%%%%%%%%%%%%%%%%%%%
\subsection*{The $I=1$, $S$-wave channel}
\nin
We have also applied FESR in the $I=1$ $S$-wave channel with the current \cite{mat.02}:
\vspace{-0.2cm}
\beq
\eta^{\Theta}_{\mbox{{\small \cite{mat.02}}}} = {1\over\sqrt{2}}\epsilon^{abc} \Big{[}  Q_{ab}^{s} 
   Q_{ce}^{s} + tQ_{ab}^{ps} 
   Q_{ce}^{ps}-(u\lrar d)\Big{]}C \bar{s}_{e}^{T}
\eeq
\vspace{-0.2cm}
where $t$ is an arbitrary mixing parameter.  To the $D\leq 5$ approximation, the analysis  gives almost the value of $M_\Theta$ in Fig. \ref{fig:
mass1} at the same approximation. This result can be interpreted as a consequence of the good realization of the $SU(2)_F$ symmetry for the $u$
and $d$ quarks. Then, we expect that the unmixed $I=1$ partners of the umnixed $I=0$ state will be around the 1.5 region
if any.
%%%%%%%%%%%%%%%%%%%%%%%%%%%%%%%%%%%%%
\section*{The $I=0$, $P$-wave channel}
\begin{figure}[hbt]
\begin{center}
\includegraphics[width=6cm]{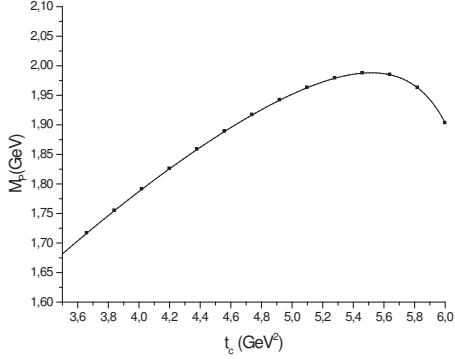}
\vspace{-.5cm}
\caption{\footnotesize $t_c$-behaviour of $M_P$ including the $D\leq 5$ condensates in the OPE. } 
\vspace{-1.cm}
\label{fig: pmass}
\end{center}
\end{figure}

\begin{figure}[hbt]
\begin{center}
\includegraphics[width=6cm]{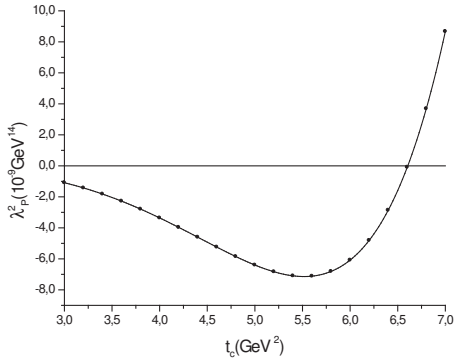}
\vspace{-.5cm}
\caption{\footnotesize $t_c$-behaviour of $\lambda^2_P\times 10^9$ GeV$^{12}$ including the $D\leq 5$ condensates in
the OPE. } 
\vspace{-1.cm}
\label{fig: plambda}
\end{center}
\end{figure}

\nin
We do a similar analysis for the $P$-wave
current given in Eqs. (\ref{RD1.eidcur}) and (\ref{eta.new}), where as we have mentioned in section 3, the contribution from Eq.
(\ref{eta.new}) vanishes to leading order in $\alpha_s$. The corresponding FESR up to
$D=5$ condensates are given below
\footnote{The inclusion of higher dimension condensates is in progress.}:
\bea
{\cal M}^P_{0,\cite{makus.04}} &=& {\frac {{\it m_s}\,{{\it t_c}}^{7}}{{361267200}{\pi }^{8}}
}-{\frac {{\la\bar{s}s\ra}\,{{ t_c}}^{6}}{5529600\pi ^{6}}}\nnb\\
&+&
\,{\frac {{\it g\la\bar{s}{\bf\sigma.G}s\ra}\,{{\it t_c}}^{5}}{614400{\pi }^{6}}-\frac {m_s \la\alpha_s
G^2\ra t_c^5}{19660800\pi^7}}
\eea
\bea
{\cal M}^P_{1,\cite{makus.04}} &=& {\frac {{\it m_s}\,{{\it t_c}}^{8}}{{412876800}{\pi }^{8}}
}-{\frac {{\la\bar{s}s\ra}\,{{ t_c}}^{7}}{6451200\pi ^{6}}}\nnb\\
&+&
\,{\frac {{\it g\la\bar{s}{\bf\sigma.G}s\ra}\,{{\it t_c}}^{6}}{737280{\pi }^{6}}-\frac {m_s \la\alpha_s
G^2\ra t_c^5}{23592960\pi^7}}
\eea
To this order, the moments present a similar $t_c$-behaviour as above (see Figs. \ref{fig: pmass} and \ref{fig: plambda}.).
%%%%%%%%%%%%%%%%%%%%
%\vspace{-.5cm}
%%%%%%%%%%%%%%%%%%%%
 Both for the mass and residue, the stability point is:
\beq
{t_c}\simeq 5.5~ {\rm GeV}^2~. 
\eeq
The corresponding resonance mass and residue are:
\bea\label{eq: pmass}
M_P&\simeq& 1.99\pm 0.19~{\rm GeV},\nnb\\
\lambda_P^2&\approx& -(0.7\sim 7.1)\times 10^{-9}~{\rm GeV}^{14}~.
\eea
The errors come mainly from the estimate of the unknown $D=7,~9$ condensates contributions inspired from the
$S$-wave channel.
%%%%%%%%%%%%%%%%%%%%
%\vspace{-.5cm}
%%%%%%%%%%%%%%%%%%%%
 One can notice that:
\\
$\b$ The value of the QCD continuum threshold at which the FESR stabilizes is much higher
than the intuitive choice used in the LSR \cite{makus.04} needed to reproduce the experimental mass of the $\Theta$.\\
$\b$ The mass value obtained for the $P$-wave resonance is $(450\pm 190)$ MeV higher than the $\Theta(1540)$ mass,
which suggests that there is a $P$-wave state different from the $\Theta(1540)$ in the region around 2 GeV, which we expect to
be discovered experimentally.\\
$\b$ The value and sign of the residue suggest that this $P$-wave state has a negative parity like the $\Theta(1540)$.
%%%%%%%%%%%%%%%%%%%%%%%%%%%%%%%%%
\section{THE $\Theta$-$K$-$N$ COUPLING}
%%%%%%%%%%%%%%%%%%%%%%%%%%%%%%%%%
\nin
For studying this coupling, we start from the three-point function:
\bea
V(p,q)=i^2\int d^4x~d^4y~e^{i(px+qy)}~\la 0\vert \eta(0)N(x)K(y)\vert 0\ra
\eea
where $K(y)\equiv (m_s+m_u)\bar s(i\gamma_5)u$ is the kaon current, while $N(x)\equiv :u(C\gamma_5)du:+b:u(Cd)\gamma_5 u:$ is the
nucleon interpolating field
\cite{HEIDEL} ($b$ being an arbitrary mixing parameter). $\eta$ is the $\Theta$ current defined in previous section. For
definiteness,  we work with the $S$-wave
current given by \cite{oka.04}. A QCD evaluation of the vertex in the chiral limit $m_s=0$ shows that the leading and
$\alpha_s$ orders perturbative and non-perturbative diagrams give zero contributions \footnote{Our result is stronger than
the one in Ref. \cite{IOFFE} which claims a non-zero $\alpha_s$ contribution.} . The result then suggests that the
$\Theta$-$K$-$N$ coupling is of the order $\alpha_s^2$ supporting the experimental observation \cite{NICOLAI} that the $\Theta (1540)$ is a
narrow state. The narrowness of a pentaquark state has been already advocated in the past, from duality arguments, where its decay into
$B\bar BB$ baryon states is dynamically favoured, implying that light pentaquark states below this threshold are naturally narrow
\cite{VENEZIA}. A narrow $S$-wave pentaquark state has been also obtained in \cite{STECH} and \cite{IOFFE} using simple chiral
symmetry arguments. In \cite{KARL} the narrowness of the $\Theta$ is due to a destructive interference between two almost degenerate
states, while in \cite{SORBA}, it is due to the flavour structure of the $\Theta$, which, after the meson formation, the residual
three-quark piece has a little overlap with the octet baryon wave-function.
%%%%%%%%%%%%%%%%%%%%%%%%%%%%%%%%%%
\section{SUMMARY AND CONCLUSIONS}
%%%%%%%%%%%%%%%%%%%%%%%%%%%%%%%%%%
\nin
$\b$ We have re-analyzed the existing LSR results in the literature. We found that due to the slow convergence of the OPE
and to the relative importance of the QCD continuum contribution into the spectral function, the minimal duality ansatz ``one
resonance + QCD continuum" is not sufficient for finding a {\it sum rule window} where the results are optimal. These
features penalize {\it all}  existing sum rule results in the literature, which then become unreliable despite the fact
that the mass predictions reproduce quite well the data. However, this {\it apparent} good prediction is due to the fact
that the {\it intuitive or arbitrary} choice of the continuum threshold $t_c$. In fact, in the LSR analysis, the mass
prediction increases with $t_c$ though it is a smooth function of $\tau$ as can be seen in Fig. \ref{fig: lsr}. \\ 
$\b$ On the contrary,
FESR has the advantage to present a good
$t_c$-stability and converges faster than the LSR, because the optimal results are obtained at higher scale $t_c\approx (2\sim
3)$ GeV$^2$ than the one of LSR $\tau^{-1}\leq 1$ GeV$^2$. \\
$\b$ Truncating the OPE at the $D=9$ condensates, at which the OPE starts to converge, we obtain the result
in Eq. (\ref{eq:smass}), for the $S$-wave state, which one can compare with the experimental candidate $\Theta(1540)$.
\\
$\b$ By truncating the OPE at $D=5$, we also find from FESR a good degeneracy between the unmixed $I=0$ and $I=1$ $S$-wave
states.\\
$\b$ Similarly, we obtain, from FESR, the  mass of the $P$-wave state of about 2 GeV in Eq. (\ref{eq: pmass}) to order $D= 5$ of
the OPE, but including the estimated effects of $D=7,9$ condensates. This mass is $(450\pm 190)$ MeV higher
than the
$\Theta(1540)$. \\
$\b$ Finally, an analysis of the
$\Theta$-$K$-$N$ coupling using vertex sum rules supports results in the literature that the $\Theta(1540)$ is a narrow
state.\\
$\b$ Our results seem to favour the case (b) discussed in \cite{JAIN} where the $\Theta$ resonance is induced in $KN$ scattering
by coupling to a confined channel. A complete program using FESR in different pentaquark channels is in progress and will be
published elsewhere.
%%%%%%%%%%%%%%%%%%%%%%%%%%%
\section*{ACKNOWLEDGEMENTS}
%%%%%%%%%%%%%%%%%%%%%%%%%%%
\nin
Communications with M. Eidemueller, R.L. Jaffe, F. Navarra, M. Nielsen, J.M. Richard, R. Rodrigues da Silva and G.C. Rossi 
are acknowledged.

\end{document}